\documentclass[aps,preprint,superscriptaddress,floatfix]{revtex4-1}

\usepackage{graphicx}
\usepackage{color}

\begin{document}


\title{Magnetotransport evidence of irreversible spin reorientation in the collinear antiferromagnetic state of underdoped
$\mathrm{Nd}_{2-x}\mathrm{Ce}_x\mathrm{CuO}_4$}


\author{A. Dorantes}
\email[]{alma.dorantes@wmi.badw.de}
\affiliation{Walther-Mei{\ss}ner-Institut, Bayerische Akademie der Wissenschaften, D-$85748$ Garching, Germany}
\affiliation{Physik-Department, Technische Universit{\"a}t M{\"u}nchen, D-$85748$ Garching, Germany}

\author{A. Alshemi}
\altaffiliation{Current address: Center for Imaging and Microscopy, Zewail city of Science and Technology. $12578$ Giza, Egypt}
\affiliation{Walther-Mei{\ss}ner-Institut, Bayerische Akademie der Wissenschaften, D-$85748$ Garching, Germany}
\affiliation{Physik-Department, Technische Universit{\"a}t M{\"u}nchen, D-$85748$ Garching, Germany}

\author{Z. Huang}
\altaffiliation{Current address: Department of Physics \& Astronomy, Rutgers University. New Brunswick, NJ $08901$, USA}
\affiliation{Walther-Mei{\ss}ner-Institut, Bayerische Akademie der Wissenschaften, D-$85748$ Garching, Germany}
\affiliation{Physik-Department, Technische Universit{\"a}t M{\"u}nchen, D-$85748$ Garching, Germany}

\author{A. Erb}
\affiliation{Walther-Mei{\ss}ner-Institut, Bayerische Akademie der Wissenschaften, D-$85748$ Garching, Germany}
\affiliation{Physik-Department, Technische Universit{\"a}t M{\"u}nchen, D-$85748$ Garching, Germany}

\author{T. Helm}
\altaffiliation{Current address: Max Planck Institute for Chemical Physics of Solids. $01187$ Dresden, Germany}
\affiliation{Walther-Mei{\ss}ner-Institut, Bayerische Akademie der Wissenschaften, D-$85748$ Garching, Germany}
\affiliation{Physik-Department, Technische Universit{\"a}t M{\"u}nchen, D-$85748$ Garching, Germany}

\author{M.V. Kartsovnik}
\email[]{mark.kartsovnik@wmi.badw.de}
\affiliation{Walther-Mei{\ss}ner-Institut, Bayerische Akademie der Wissenschaften, D-$85748$ Garching, Germany}


\date{\today}

\begin{abstract}
We make use of the strong spin-charge coupling in the electron-doped cuprate $\mathrm{Nd}_{2-x}\mathrm{Ce}_x\mathrm{CuO}_4$
to probe changes in its spin system via magnetotransport measurements.
We present a detailed study of the out-of-plane magnetoresistance in underdoped single crystals of this compound,
including the nonsuperconducting, $0.05\,\leq x\,\leq 0.115$, and superconducting, $0.12\,\leq x\,\leq 0.13$, compositions.
Special focus is put on the dependence of the magnetoresistance on the field orientation in the plane of the CuO$_2$ layers.
In addition to the kink at the field-induced transition between the noncollinear and collinear antiferromagnetic configurations,
a sharp irreversible feature is found in the angle-dependent magnetoresistance of all samples in the high-field regime,
at field orientations around the Cu--O--Cu direction.
The obtained behavior can be explained in terms of field-induced reorientation of Cu$^{2+}$ spins within the collinear
antiferromagnetic state. It is, therefore, considered as an unambiguous indication of the long-range magnetic order.
\end{abstract}

\maketitle

\section{Introduction}
\label{intro}
The parent compounds of the high-temperature cuprate superconductors are known to be antiferromagnetic (AF) Mott insulators and become metallic upon charge doping. Unlike the hole-doped cuprates, in which the long-range AF order is suppressed already at low doping, the electron-doped compounds remain AF up to at least the superconducting (SC) doping range \cite{Armitage2010}.
However, whether the AF order coexists with superconductivity, and if yes, to which extent, remains a matter of controversy.
A number of neutron scattering and muon spin relaxation studies sensitive to the magnetic structure suggest a static or quasistatic AF order in the electron-doped cuprates over a large part of the SC doping range \cite{kang03,Yamada2003,fuji08,wata01,mang04}, others set the limit of ordering at the lower border of superconductivity \cite{motoyama_2007,Mang2004,Fujita2003,Luke1990,saad15}. Most of angle-resolved photoemission spectroscopy (ARPES) experiments indicate a Fermi surface reconstruction ascribed to AF ordering up to optimal SC doping \cite{armitage_doping_2002,matsui_2007,sant11,kyun04}, whereas in a very recent work \cite{song17} only short-range AF fluctuations have been detected throughout the SC region. Furthermore, magnetic quantum oscillations show that the Fermi surface is still reconstructed up to the upper border of the SC doping range \cite{helm10,Kartsovnik2011,helm15}. However, it is not clear whether this reconstruction is caused by the AF order or by the recently detected charge-density modulation \cite{lee14,silv15}.

Classical magnetotransport has extensively been used for exploring the electronic state of electron-doped cuprates and in particular for searching for manifestations of magnetic ordering \cite{Lavrov2004,Chen2005,Li2005,Fournier2004,Ponomarev2005,Wu2008,daga04,daga05,li07b,jenk10,Yu2007,Jovanovic2010,Kartsovnik2011,helm15}. An obvious advantage of transport measurements in comparison to surface sensitive methods, such as scanning tunneling microscopy or ARPES is that they probe bulk properties throughout the sample. Unlike neutron or magnetization techniques, they can be done on very small samples, which are easier to obtain with the required crystal quality. Furthermore, being sensitive specifically to the conducting system, they do not suffer from the presence of spurious insulating phases caused, e.g., by postgrowth annealing \cite{Kang2007, Mang2004}. On the other hand, the charge or heat transport is of course only an indirect probe of the magnetic state.
Fortunately, some prominent transport features can be shown to directly correlate with transformations in the AF spin structure and as such can serve as unambiguous indications of magnetic ordering in these compounds. A remarkable example is the sharp step in magnetoresistance found in Pr$_{1.3-x}$La$_{0.7}$Ce$_x$CuO$_4$ \cite{Lavrov2004}, $\mathrm{Nd}_{2-x}\mathrm{Ce}_x\mathrm{CuO}_4$ (NCCO) \cite{Li2005,Chen2005} at low doping and in as-grown (non-SC) $\mathrm{Pr}_{1.85}\mathrm{Ce}_{0.15}\mathrm{CuO}_4$ \cite{Fournier2004} crystals. This step is caused by the phase transition between the noncollinear orientation of the Cu$^{2+}$ spins in adjacent layers, which is stable at zero magnetic field, to the high-field collinear AF state \cite{Petitgrand1999,Plakhty2003}.

Here we present another magnetotransport feature which can be considered as a fingerprint of the AF state in the electron-doped cuprates. We have carried out systematic studies of the interlayer magnetoresistance of underdoped NCCO single crystals with the Ce concentrations near the border of the SC range of the phase diagram. Besides the sharp step in the field- and angle-dependent magnetoresistance, corresponding to the transition between the collinear and noncollinear states, we have observed a prominent hysteretic feature in the high-field, collinear state at field orientations around the $[100]$ direction. While this feature is similar to the hysteretic anomaly reported earlier for strongly underdoped NCCO and attributed to the ordering of Nd$^{3+}$ spins \cite{Wu2008}, it is found to persist at temperatures strongly exceeding the N\'{e}el temperature of the Nd$^{3+}$ system. We propose a qualitative explanation of this anomaly based on the model of the field-dependent orientation of Cu$^{2+}$ spins in the collinear state. Interestingly, the high-field hysteretic behavior has been found not only on the non-SC samples but also on the SC samples . This result points to the coexistence of superconductivity and antiferromagnetism, at least, near the lower edge of the SC doping range.

\section{Experiment}
\label{exp}
Single crystals of NCCO with Ce concentration in the range $x= 0.05 - 0.13$ were grown with the traveling solvent floating zone method, as described earlier \cite{Lambacher2010}. From an as-grown crystal rod, samples with dimensions of $\approx 0.3\times 0.3\times 1\,\mathrm{mm}^{3}$ were cut out. The longest dimension corresponded to the $[001]$ ($c$-axis) crystallographic direction. The samples were annealed in the argon atmosphere according to the dopant concentration \cite{Lambacher2010}: $850^{\circ}\mathrm{C}$ for the $x = 0.05$ sample, $900^{\circ}\mathrm{C}$ for $x = 0.09$ and $0.10$ samples, $910^{\circ}\mathrm{C}$ for the $x =  0.115$, $0.12$ and $0.125$ samples, and $935^{\circ}\mathrm{C}$ for $x =  0.13$. Samples with $x\,\leq 0.115$ were annealed for $20$ hours and samples with $x\,\geq 0.12$ for $40$ hours.
The samples $0.05\,\leq x\,\leq\,0.115$ were non-superconducting (non-SC). The resistivity of the $x = 0.115$ sample showed a tiny, $\approx 3\%$ downturn below $T\,=\,7.7\,\mathrm{K}$. Although this indicates the presence of a very small fraction of superconducting volume, we refer to this sample as non-SC. The samples with $x = 0.12, 0.125$, and $0.13$ had a full SC transition according to resistivity and zero-field-cooling SQUID magnetization measurements.

The electrical contacts were prepared for 4-probe measurements of the interlayer resistance, using conducting Ag-based epoxy as described in Ref. \cite{Kartsovnik2011}.
The contacted samples were mounted on a rotatable platform so that the rotation axis was parallel to the $[001]$ direction.
The platform was placed in the center of a superconducting solenoid, with the rotation axis perpendicular to the solenoid axis.
Thus, the external magnetic field was always directed perpendicular to the current and parallel to the CuO$_2$ layers, and a continuous \textit{in situ} rotation of the samples with respect to the field direction was possible.
The angle-dependent magnetoresistance was measured as a function of the azimuthal angle $\varphi$ between the $[100]$ axis and the field direction, at different fixed field strengths $B \leq 15$\,T, in the temperature range 1.4 to 115\,K.
Additionally, continuous field sweeps were done for the field orientations along the Cu--O--Cu (crystallographic $[100]$ or $[010]$ axis), Cu--Cu ($[110]/[1\bar{1}0]$ axis), and some intermediate directions within the layer plane.

In accordance with the tetragonal crystal symmetry of NCCO and the axial symmetry with respect to the applied current direction, the AMR showed a $90^\circ$ periodicity \cite{Comment_4fold}. Therefore, in what follows, our
discussions regarding the directions $[100]$ and $[110]$ are also valid for the equivalent directions $[010]$ and $[1\bar{1}0]$, respectively.

\section{Results}
\label{results}

\begin{figure}
  \centering
  \includegraphics[width=0.45\textwidth]{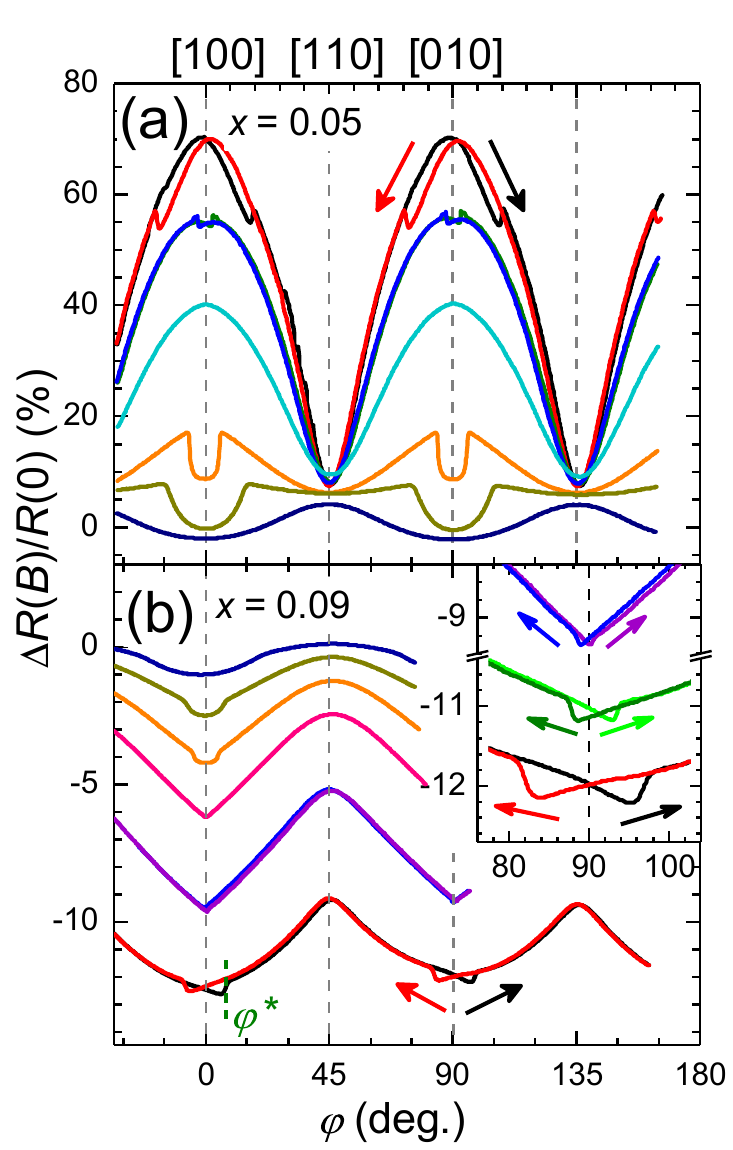}
  \caption{(Color online) AMR of non-SC samples for field rotations in the plane of CuO$_2$ layers, at different field strengths, $T=1.4$\,K.
  (a) $x = 0.05$; the field is (from bottom to top): $B = 1, 1.5, 3, 5, 8$, and $15\,\mathrm{T}$; (b) $x = 0.09$; the field is (top to bottom): $B = 1, 2, 3, 4, 6, $ and $14\,\mathrm{T}$. Up and down $\varphi$-sweeps, as pointed by the arrows of the respective colors, are shown for the high-field AMR, revealing a hysteresis around $\varphi =0^\circ$ and $90^\circ$. The inset is a close-up of the hysteresis around $\varphi=90^{\circ}$ at  $B = 6, 10,$ and $14\,\mathrm{T}$ for the $x=0.09$ sample.}
\label{fig:AMR_N5_N9_lowT}
\end{figure}
Figure\,\ref{fig:AMR_N5_N9_lowT} shows the angle-dependent magnetoresistance (AMR) of two underdoped non-SC samples, $x=0.05$ and $0.09$, at $T\,=\,1.4\,\mathrm{K}$ and different magnetic fields $1\,\mathrm{T}\,\leq\,B\,\leq 15\,\mathrm{T}$. Here $\Delta R(B)/R(0) \equiv R(B) - R(B=0)/R(B=0)$. Below $6$\,T the AMR is fully reversible. The overall shape and magnitude are consistent with the field-dependent magnetoresistance patterns presented for $\mathbf{B} \| [100]$ and $\mathbf{B} \| [110]$ in the Appendix. At $B=1$\,T, the resistance of both samples increases from a minimum at $\mathbf{B}\|[100]$, corresponding to the noncollinear spin arrangement to a maximum at $\mathbf{B}\|[110]$, where the high-field collinear AF state is already more stable at this field strength \cite{Lavrov2004,Chen2005,Wu2008}. With increasing $B$, the angular range, in which the noncollinear state with a relatively low magnetoresistance exists, becomes more narrow.
The transition to the collinear state, seen for both samples as a step in magnetoresistance, gradually shifts towards $[100]$. Eventually the collinear state is set in the entire angular range when the field exceeds the highest critical value $B_{c,\mathrm{max}} = B_c(\varphi=0^\circ) \approx 3.8$\,T, see Appendix. Our other non-SC samples, with $0.09 \leq x \leq 0.115$, showed the AMR similar to that in Fig.\,\ref{fig:AMR_N5_N9_lowT}(b).


At fields $B\geq\,6\,\mathrm{T}$, a new feature emerges in the AMR around the $[100]$ direction. Taking, for example, the 14\,T curve for $x=0.09$ in Fig.\,\ref{fig:AMR_N5_N9_lowT}(b), the resistance minimum shifts from the exact [100] position. As the angle increases from negative values, passing through $0^\circ$, the resistance continues decreasing until a critical angle $\varphi^*\approx 7^\circ$, at which it sharply increases. The same jump is observed upon decreasing $\varphi$ from the positive side, through $0^\circ$, at $-\varphi^*$. Thus, the AMR exhibits a hysteresis in the angular range $\Delta\varphi \approx 2\varphi^{*}$ around the $[100]/[010]$ directions. Beyond this interval the angular dependence is fully reversible. The width of the hysteresis depends on the field strength, see the inset in Fig.\,\ref{fig:AMR_N5_N9_lowT}(b). It increases from $\Delta\varphi \approx 2^\circ$ at $B=6$\,T to $\approx 15^\circ$ at $14$\,T. The samples with $x = 0.10$ and  $0.115$ show qualitatively the same behavior. Moreover, despite the opposite overall anisotropy, the high-field AMR of the lower doped sample displays a very similar hysteresis around the $[100]/[010]$ directions, see Fig.\,\ref{fig:AMR_N5_N9_lowT}(a). It also resembles the sharp hysteretic feature reported for strongly underdoped NCCO crystals, with $x = 0.025$ and 0.033 at fields $\geq 10$\,T \cite{Wu2008}. This behavior strikingly differs from the normal metallic magnetoresistance and is most likely caused by coupling of the charge transport to the AF ordered spin system. As will be discussed in Section\,\ref{discuss}, it can be explained qualitatively by a field-induced rearrangement of antiferromagnetically ordered Cu$^{2+}$ spins.

\begin{figure}[b]
  \centering
  \includegraphics[width=0.45\textwidth]{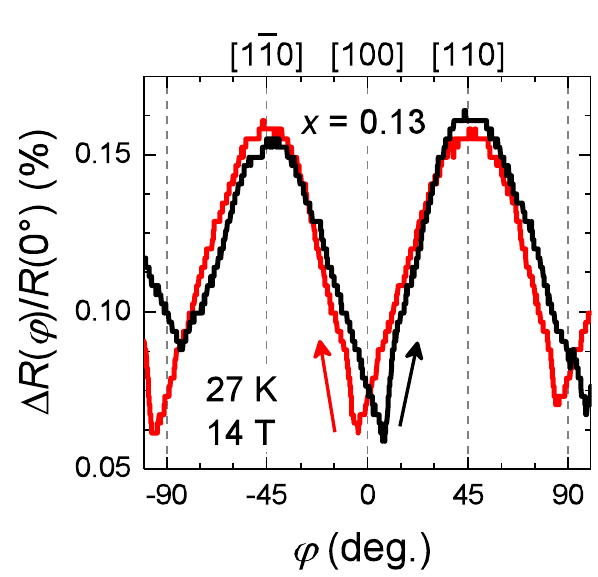}
  \caption{(Color online) AMR of a SC sample, $x = 0.13$, at $T > T_{c\mathrm{,o}}$.
  Here the magnetoresistance is defined as $\Delta R(\varphi)/R(0^{\circ}) = R(B,\varphi) - R(B,\varphi= 0^{\circ})/R(B,\varphi= 0^{\circ})$.
  The arrows show the angular sweep directions for the curves of the respective color.}
\label{fig:AMR_N13}
\end{figure}
Interestingly, a similar behavior has been found in the normal state of the underdoped SC crystals with the Ce concentrations $x = 0.12, 0.125$, and $0.13$. Due to the very high upper critical field along the layers, superconductivity in these samples could not be fully suppressed by our maximum field, 15\,T, even at temperatures $\sim 1-2$\,K below the SC onset temperature $T_{c\mathrm{,o}}$. Moreover, even a minor, $< 1^\circ$, misalignment from the exactly inplane field orientation had a strong effect on the mixed-state resistivity. As a result, the shape of the $R(\varphi)$ curves below $T_{c\mathrm{,o}}$ was mainly governed by the variation of the tiny out-of-plane field component, making it impossible to detect weak normal-state magnetoresistance features. Therefore our studies were focused on temperatures above $T_{c\mathrm{,o}}$.

As a typical example, angular up- and down-sweeps recorded for the $x = 0.13$ sample at $T=27$\,K, $B=14$\,T are shown in Fig.\,\ref{fig:AMR_N13}.
Like in the case of the non-SC samples, the hysteresis and accompanying step-like feature are clearly observed and the width of the hysteresis grows with increasing magnetic field. Qualitatively the same behavior has been found for all the other SC samples studied.

\begin{figure}[b]
  \centering
  \includegraphics[width=0.45\textwidth]{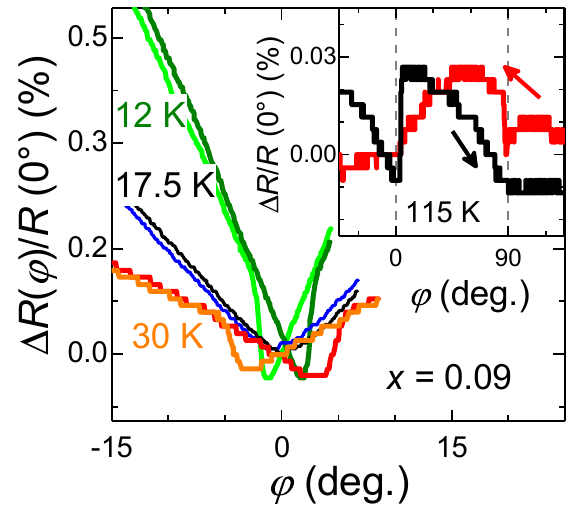}
  \caption{(Color online) Close-up of the hysteresis in the AMR of the $x = 0.09$ sample at $B=14\,\mathrm{T}$ and different temperatures. The inset shows, in a different scale, the AMR of the same sample at $T = 115\,\mathrm{K}$; the black and red arrows indicate the angular sweep directions for the curves of the respective color. The step in the magnetoresistance can bee seen near $\varphi = 0^\circ$ and $90^\circ$.}
\label{fig:AMR_N9_highT}
\end{figure}
Figure \ref{fig:AMR_N9_highT} illustrates a nonmonotonic variation of the hysteretic step-like feature with temperature, at $B=14\,\mathrm{T}$ for $x=0.09$. At 12\,K it is similar to that at 1.4\,K [see the inset in Fig.\,\ref{fig:AMR_N5_N9_lowT}(b)], although the height of the resistance step is about 2 times smaller and the hysteresis width is reduced to $\Delta \varphi \approx 3^\circ$. At 17.5\,K the feature seems to completely vanish, but it reappears at higher temperatures. For example, at 30\,K the hysteresis is even broader than at 12\,K. With increasing temperature the size of the resistance step continuously decreases together with the overall AMR amplitude; however it can be traced as long as the angular variation of the magnetoresistance is reliably measured. The inset in Fig.\,\ref{fig:AMR_N9_highT} shows an example of angular sweeps up and down at $T=115$\,K.
At this temperature the AMR amplitude, $\sim 3\times 10^{-4}$ of the total resistance, is already at the border of the experimental accuracy. The signal noise and temperature fluctuations lead to an apparent breakdown of the $90^{\circ}$ periodicity. However, the irreversible steps near the $[100]/[010]$ directions can still be resolved and have been reproduced in several sweeps done on the present sample as well as on other samples used in the experiment. We note that a similar sharp step can be seen in the AMR of a low-doped, $x=0.025$, NCCO crystal at $T=100$\,K, $B=12$\,T reported by Chen et al. \cite{Chen2005}. Although only one sweep direction was shown in that work and no comment on hysteresis was made, it seems to be directly related to the feature discussed here.


\begin{figure}
  \centering
  \includegraphics[width=0.45\textwidth]{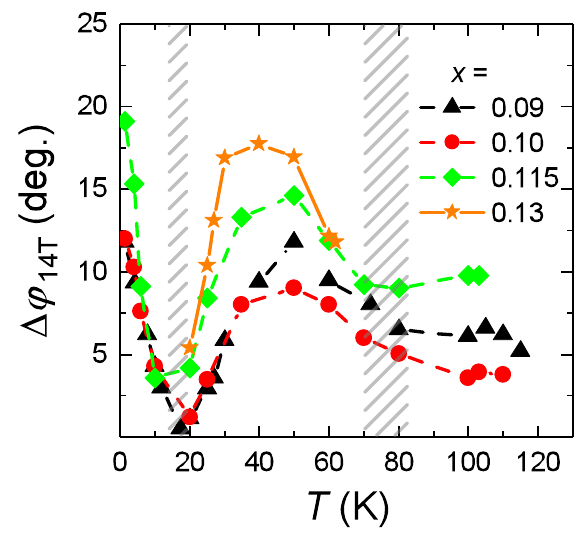}
  \caption{(Color online) The width of the hysteresis in the AMR recorded at $B=14$\,T, as a function of temperature for three non-SC samples, $x\leq 0.115$, and one SC sample, $x=0.13$.}
\label{fig:Hyst}
\end{figure}
The temperature dependence of the hysteresis width at 14\,T, $\Delta\varphi_\mathrm{14T}(T)$, is shown in Fig.\,\ref{fig:Hyst}. All samples display qualitatively the same behavior. For the non-SC crystals the hysteresis is widest at the lowest temperature, $1.4\,\mathrm{K}$. It first narrows with increasing $T$ and shows a minimum, possibly even vanishes at a temperature slightly below $20\,\mathrm{K}$. Above $20\,\mathrm{K}$, the hysteresis reappears, passes through a broad maximum around $\sim50\,\mathrm{K}$, and eventually saturates at a level of 5-$10^\circ$ above $70$\,K.

For the SC samples the temperature interval for the observation of the hysteretic feature is more narrow. On the lower side it is limited by the SC onset, as mentioned above. On the other hand, the weakened overall AMR magnitude restricts the observation of the hysteresis to temperatures below $\sim 60-70$\,K. Nevertheless, within the available $T$ range, the SC samples exhibit the same trend as the non-SC ones. For example, for the $x=0.13$ sample shown in Fig.\,\ref{fig:Hyst} $\Delta \varphi$ grows with increasing the temperature above 20\,K and passes through a broad maximum. The only quantitative difference is that the temperature of the maximum, $\simeq 40$\,K, seems to be somewhat lower than for the lower doped, non-SC samples.
%
%
%

\begin{figure}
  \centering
  \includegraphics[width=0.45\textwidth]{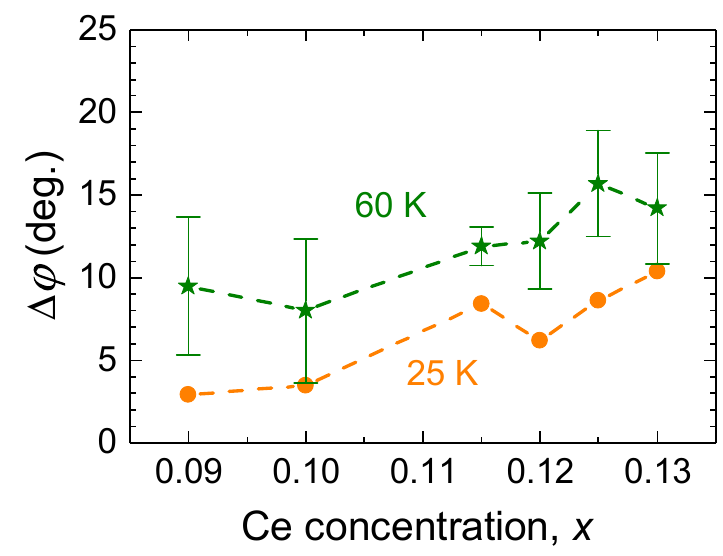}
  \caption{(Color online) The width of the hysteresis in the 14\,T AMR patterns obtained at $T= 25$ and 60\,K for different SC and non-SC samples, plotted against the nominal doping level $x$.}
\label{fig:Hyst_25K_60K}
\end{figure}

In Fig.\,\ref{fig:Hyst_25K_60K} we compare the hysteresis width obtained for different $x$ at $T= 25$ and $60$\,K.
There is significant scattering. On the one hand, this is caused by a relatively large error bar, especially at the higher temperature, due to the weakness of the feature. On the other hand, one has to keep in mind a possible dependence of the hysteresis on sample quality. Nevertheless one can trace a general trend of increasing $\Delta \varphi$ with increasing the doping level.

In earlier magnetotransport studies of lightly doped NCCO, with $x=0.025$, sharp changes in the AMR shape have been detected at $T\approx 70$\,K \cite{Chen2005} and $\approx 30$\,K \cite{Chen2005,Li2005}. They were attributed to successive transitions between AF phases I and II, and III, characterized by different mutual orientations of Cu$^{2+}$ and Nd$^{3+}$ spins \cite{Matsuda1990,skan93,Sachidanandam1997,Li2005}. Our experiments on stronger doped samples, $x \geq 0.09$, do not show such sharp changes of the overall out-of-plane AMR. However, the nonmonotonic $\Delta\varphi(T)$ dependence in Fig.\,\ref{fig:Hyst} suggests the presence of three distinct temperature regions, schematically separated by the hatched boxes in the Figure. By analogy with the undoped Nd$_2$CuO$_4$  and lightly doped NCCO samples, these three regions might be associated with the different AF spin configurations. Of course this suggestion has to be verified by direct investigations of the magnetic structure in this doping range.
We note that the lower critical temperature in Fig.\,\ref{fig:Hyst}, $\lesssim 20$\,K, is considerably lower than the temperature of the transition between the phases II and III, $T_\mathrm{N,2}\approx 30$\,K, found for $x=0$ and 0.025 \cite{Chen2005,Li2005}. This difference is not very surprising, taking into account possible modification of magnetic properties when approaching the SC doping range. Indeed, this temperature seems to be the lowest for the highest doping, $x=0.115$, which is already very close to the SC region of the phase diagram.

\begin{figure}[b]
  \centering
  \includegraphics[width=0.45\textwidth]{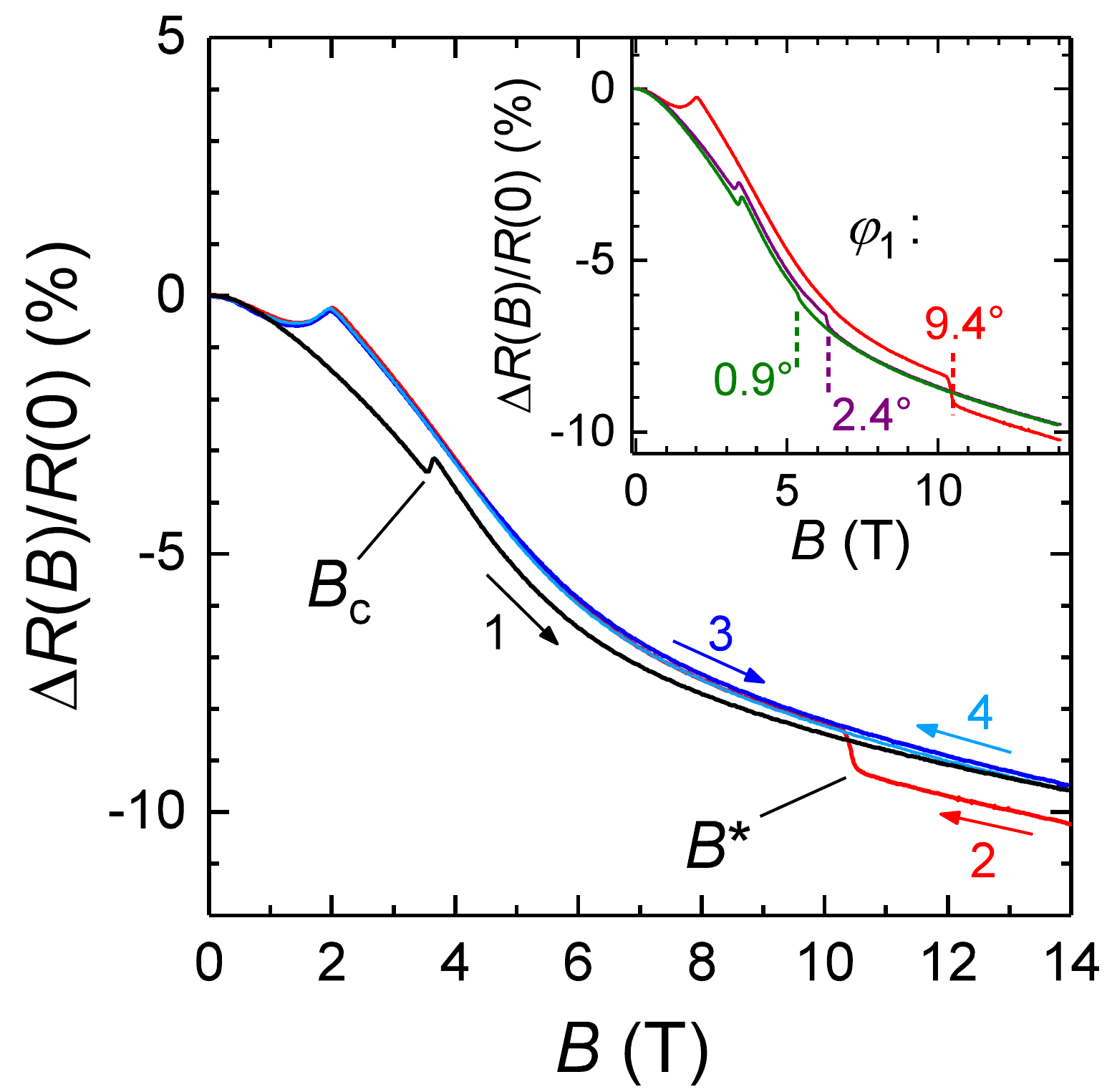}
  \caption{(Color online) Consecutive magnetoresistance field-sweeps performed on a $x = 0.10$ sample at $T=1.4$\,K, at different orientations near the $[100]$ direction. Curve 1 in the main panel was recorded at $\varphi_0 = -0.6^\circ$ and the other curves at $\varphi_1 = 9.4^\circ$. The colored arrows indicate the directions of the corresponding field sweeps, see text. Inset: down-sweeps at different angles $\varphi_1$, each done after an up-sweep at $\varphi_0=-0.6^\circ$. Vertical dashed lines point to the respective positions $B^*$ of the irreversible step-like feature.}
\label{fig:N10_Bstar}
\end{figure}
The hysteretic behavior of the AMR gives rise to an interesting memory effect in the field-dependent magnetoresistance. In Fig.\,\ref{fig:N10_Bstar} we show a sequence of low-$T$ field sweeps on a $x=0.10$ sample illustrating this effect.
First, the sample is cooled down to 1.4\,K at $B=0$ a field sweep is done up to 14\,T at $\varphi_0 =-0.6^\circ$, curve 1 in Fig.\,\ref{fig:N10_Bstar}.
A sharp step at $B_c \approx 3.6$\,T indicates the transition into the field-induced collinear AF state (see Appendix) \cite{Comment_sample10}.
Next, at the constant field, the angle is changed to positive $\varphi_1=9.4^\circ$ and the field is driven down at this angle. Expectedly, since the field is now more strongly tilted from the $[100]$ direction, the collinear-noncollinear transition is shifted to a lower field. But what is new in the down-sweep is, that the field-dependent magnetoresistance exhibits a step in the high-field region, at $B^* \approx 10.5$\,T, see curve 2 in Fig.\,\ref{fig:N10_Bstar}. Subsequent field sweeps up (curve 3) and down (curve 4) done at the same orientation show a fully reversible behavior with the kink at $B_c$ but without the high-field step. To reproduce the step, one has to apply a high enough field and turn it, passing through the $[100]$ direction,
to an angle $\varphi_1$, such that $|\varphi_1| < \varphi^*$ ($\varphi^*$ is the half-width of the hysteresis in the corresponding constant-field AMR pattern).
Then the high-field step will be observed upon decreasing the field.
It is also sufficient to apply the field exactly along $[100]$ and then turn it to $\varphi_1$ before sweeping the field down. However, once the angle $\varphi_0$ of the initial up-sweep is on the same side from $[100]$ as $\varphi_1$, the down-sweep becomes fully reversible with no feature in the high-field state.
Following the described sequence, we were able to reproduce the high-field step-like feature in the field-dependent magnetoresistance on all the samples studied, on the non-SC as well as on the SC ones.

The position of the step in the field down-sweep depends on the field orientation, as shown in the inset in Fig.\,\ref{fig:N10_Bstar}: the smaller $\varphi_1$, the lower the characteristic field $B^*$, at which the magnetoresistance relaxes back to the reversible behavior. Simultaneously the height of the magnetoresistance step at $B^*$ decreases. This is of course consistent with the field dependence of the AMR hysteresis loop in Fig.\,\ref{fig:AMR_N5_N9_lowT}, which becomes narrower and smaller in magnitude at lowering the field.

\section{Discussion}
\label{discuss}
As follows from the data in Fig.\,\ref{fig:AMR_N5_N9_lowT}, the angular dependence of the critical field of the transition between the noncollinear and collinear AF states results in a sharp feature in the AMR at the intermediate field range, between $B_{c\mathrm{,min}}=B_c(\varphi=45^\circ)$ and $B_{c\mathrm{,max}}=B_c(\varphi=0^\circ)$. The high-field hysteretic anomaly occurs entirely in the collinear state and must be caused by some discontinuous change of the spin structure within this state. Wu et al. \cite{Wu2008} ascribed a similar anomaly found in low-doped NCCO to the ordering of Nd$^{3+}$ spins. While the temperatures reported in that work, $T\leq 5$\,K, were indeed not far away from the N\'{e}el temperature of the Nd$^{3+}$ subsystem, we observe this behavior up to much higher temperatures, above 100\,K. Therefore, its origin should rather be associated with the Cu$^{2+}$ spins, which remain ordered at these temperatures \cite{Matsuda1990,mang04}.

\begin{figure}
  \centering
  \includegraphics[width=0.45\textwidth]{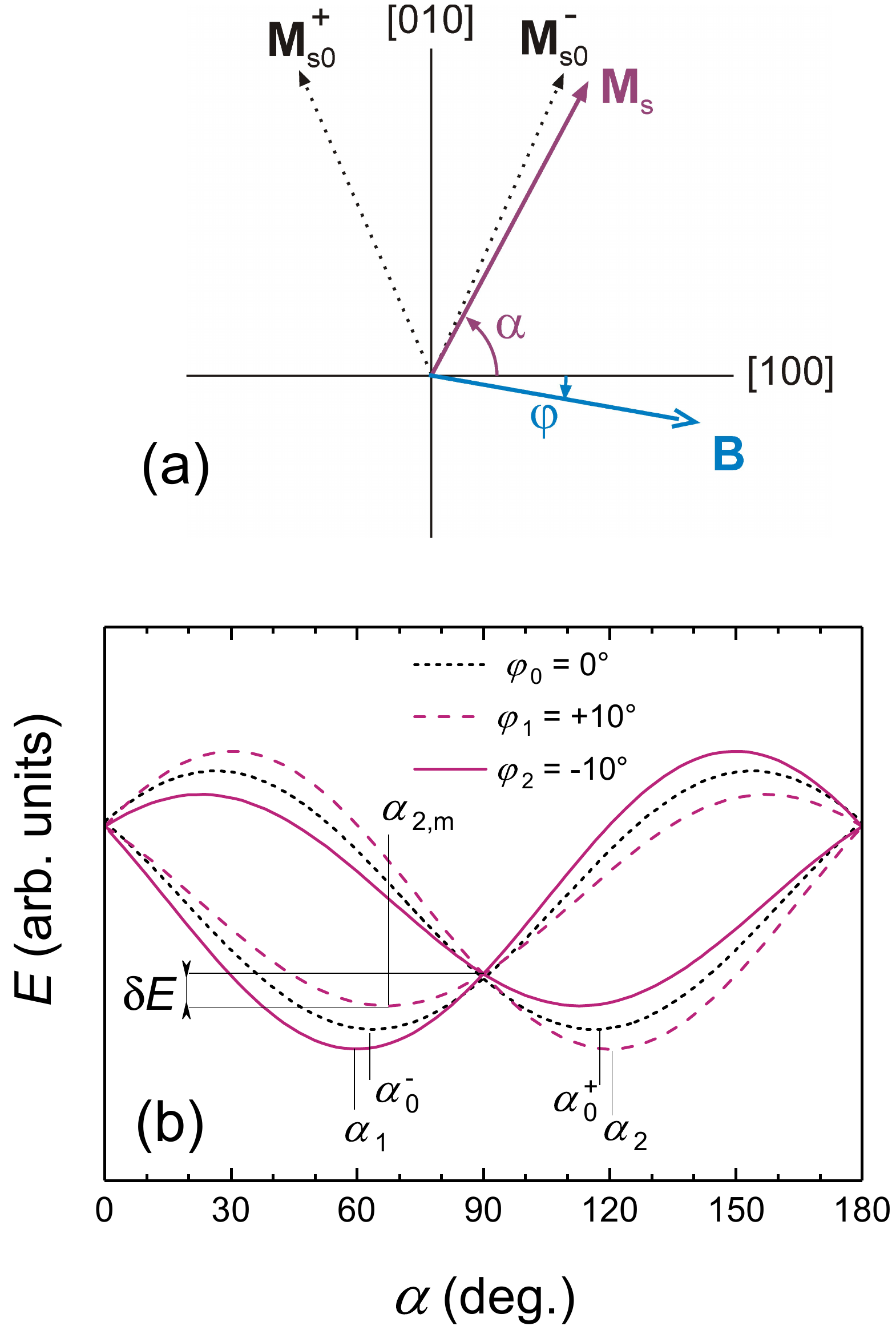}
  \caption{(Color online) (a) Schematic diagram of the equilibrium orientation of the Cu$^{2+}$ staggered moment $\mathbf{M}_{\mathrm{s}}$ in the field $B=14$\,T applied at angle $\varphi_1 = -10^\circ$ from the $[100]$ direction, obtained from Eq.\,(\ref{energy}); $\alpha$ is the angle between the staggered magnetization vector $\mathbf{M}_{\mathrm{s}}$ and $[100]$. Dotted arrows show two equilibrium directions of $\mathbf{M}_{\mathrm{s}}$ for $\mathbf{B}\| [100]$. (b) Dependence of magnetic energy on the angle $\alpha$, according to Eq.\,(\ref{energy}), for the field $B=14$\,T aligned parallel to the $[100]$ axis or at an angle of $\pm10^\circ$ from this direction. At $\varphi=0^\circ$ the state is doubly degenerate with the equilibrium angles $\alpha^{\pm}_0= 90^\circ \pm 26.3^\circ$. In a tilted field there is only one stable state with $\alpha_1$ and $\alpha_2$ for $\varphi_1$ and $\varphi_2$, respectively. $\delta E$ is the energy barrier for a transition from the metastable state at $\alpha_{2\mathrm{,m}}$ to the stable state at $\alpha_2$.
  }
\label{fig:Ms_B_Ealfa}
\end{figure}

Let us consider the dependence of the orientation of Cu$^{2+}$ spins on the field direction of a strong inplane magnetic field. This has been investigated both theoretically and experimentally, by neutron scattering, on the undoped Nd$_2$CuO$_4$ and the sister compound Pr$_2$CuO$_4$ \cite{Plakhty2003,Petitgrand1999}. We are now particularly interested in the high-field, collinear state. An important observation made by Plakhty et al. \cite{Plakhty2003} is that even in this state the spins are not perpendicular to the field, unless the latter is applied exactly along $[110]$. The equilibrium direction of the Cu$^{2+}$ staggered moment $\mathbf{M}_{\mathrm{s}}$ in the high-field collinear state is mainly determined by the balance of the contributions of the pseudodipolar and Zeeman terms to the energy, which can be expressed as \cite{Plakhty2003,Petitgrand1999}:
\begin{equation}
E \approx E_0\left[\pm G\sin 2\alpha - 2K^2\sin^2\left(\varphi - \alpha\right)\right]\,,
\label{energy-tot}
\end{equation}
where $E_0$ is determined by the inplane exchange interaction, $G=(\Omega_{\mathrm{opt}}/\Delta_0)^2$ with $\Delta_0$ being the in-plane spin-wave gap and $\Omega_{\mathrm{opt}}$ the splitting of the in-plane spin-wave spectrum caused by the interplane pseudodipolar interaction, and $K=g\mu_\mathbf{B}B/\Delta_0$ characterizes the Zeeman splitting. The angles $\varphi$ [defined in the interval $(-\frac{\pi}{2},\frac{\pi}{2})$] and $\alpha$ give the directions of the field $\mathbf{B}$ and staggered moment $\mathbf{M}_\mathrm{s}$, respectively, as shown in Fig.\,\ref{fig:Ms_B_Ealfa}(a). We put the sign ``$\pm$'' in front of the first term in Eq.\,(\ref{energy-tot}) to take into account the tetragonal symmetry of the system.
In particular, for $\varphi =0^\circ$ it yields two equilibrium directions, $\mathbf{M}_{\mathrm{s},0}^+$ and $\mathbf{M}_{\mathrm{s},0}^-$ (the other two solutions with the staggered magnetization vectors turned by $180^\circ$ are of course physically the same).
Knowing the critical fields $B_{c\mathrm{,max}}$ and $B_{c\mathrm{,min}}$ for the transition between the noncollinear and collinear states at $\mathbf{B}\|[100]$ and $\mathbf{B}\|[110]$, respectively, we can estimate the coefficients in Eq.\,(\ref{energy-tot}) \cite{Petitgrand1999}: $K = B/B_{c\mathrm{,min}}$ and $G \approx \left(B_{c\mathrm{,max}}/B_{c\mathrm{,min}}\right)^4$. Substituting the typical values $B_{c\mathrm{,max}} = 4$\,T and $B_{c\mathrm{,min}} =1$\,T, we rewrite the dependence of the energy on the angle $\alpha$ as:

\begin{equation}
E \propto \left[\pm \sin 2\alpha - 0.0078\,\mathrm{T}^{-2} \cdot B^2\sin^2\left(\varphi - \alpha\right)\right].
\label{energy}
\end{equation}

The two branches of this dependence (corresponding, respectively, to different signs of the first term) are shown in Fig.\,\ref{fig:Ms_B_Ealfa}(a) for a field of $B=14$\,T directed at $\varphi = 0^\circ$ and $10^\circ$.
As mentioned above, the energy exhibits two equal minima for $\varphi = 0^\circ$, at $\alpha_0^- = 63.7^\circ$ and $\alpha_0^+ = 116.3^\circ$. Hence, if the field is increased from zero to 14\,T exactly parallel to $[100]$, in the collinear state one should expect a formation of domains with staggered moments directed as shown by dotted arrows in Fig.\,\ref{fig:Ms_B_Ealfa}(a). This degeneracy is lifted once the field is tilted from $[100]$.
For example, for a field of 14\,T applied at angle $\varphi_{1} =-10^\circ$ the energy in Eq.\,(\ref{energy}) exhibits only one global minimum, at $\alpha_{1}= 59.8^\circ $. Consequently, if the field is ramped up from zero at a constant $\varphi$, only one direction of $\mathbf{M}_{\mathrm{s}}$, smoothly dependent on the field strength, should be realized in the collinear state, see Fig.\,\ref{fig:Ms_B_Ealfa}(a). At the critical field $B_c(\varphi)$ the spin alignment is almost parallel to $[110]$ and asymptotically approaches the direction perpendicular to the field at $B \rightarrow \infty$ \cite{Plakhty2003}.

The situation becomes more complicated, if the sample is turned at a high field, passing through the direction $\mathbf{B} \| [100]$, for example at turning from $\varphi_1=-10^\circ$ to $\varphi_2=10^\circ$.
The global minimum of the energy is gradually shifting from $\alpha_1$ towards $\alpha_0^-$, see Fig.\,\ref{fig:Ms_B_Ealfa}(b), as the field angle approaches $0^\circ$ from the negative side. When $\varphi$ crosses
zero, the global minimum jumps discontinuously from $\alpha < \alpha_0^-$ to $\alpha > \alpha_0^+$ and then goes on deepening and shifting to $\alpha_2$ as $\varphi$ reaches $\varphi_2$.
However, the ``$-$'' branch of Eq.\,(\ref{energy}) still has a local minimum at angle $\alpha_{2\mathrm{,m}}<90^\circ$. If the energy barrier $\delta E$ between the two minima (Fig.\,\ref{fig:Ms_B_Ealfa}(b)) is high enough, most of the spins still remain in the metastable state with $\alpha_{2\mathrm{,m}}$. As the field is further tilted away from $[100]$, the barrier is reduced and at a critical angle
$\varphi^*$ the spins relax to the equilibrium orientation at $\alpha > 90^\circ$. Obviously, the same evolution is expected of a rotation of the field in the opposite direction. Due to the strong spin-charge coupling in our system, the described field-induced spin reorientation effects should be seen in the magnetoresistance, which
is qualitatively consistent with the hysteretic AMR behavior observed in our experiment.

So far we were only considering the Cu$^{2+}$ spin system. As to the Nd$^{3+}$ spins, they appear to be indirectly involved in the observed phenomena. This is evidenced, for example, by an anomaly in low-temperature angle-dependent magnetization, strongly dominated by Nd$^{3+}$, which has been found in magnetic fields slightly tilted from the $[100]$ axis \cite{Wu2008,bazh04}. Due to the exchange interaction with Cu$^{2+}$, the orientation of the paramagnetic Nd$^{3+}$ spins is sensitive to the changes described above. This, in  particular, may be a reason for an enhancement of the hysteretic feature at low temperatures, at which the magnetic susceptibility rapidly grows \cite{Wu2008}. However, the basic origin of this behavior lies in the field-induced rearrangement of the antiferromagnetically ordered Cu$^{2+}$ spins.

While the presented model explains, at least qualitatively, the existence of the hysteresis in the angular sweeps, its dependence on the field strength remains an open question. According to Eq.\,(\ref{energy}), an increase of the field should lead to a shift of the energy minima closer to $90^\circ$ and to a reduction of the energy barrier between the metastable and stable states. Therefore one would expect that the hysteresis width $\Delta \varphi$ reduce at a higher $B$. This apparently contradicts the data in the inset of Fig.\,\ref{fig:AMR_N5_N9_lowT}(b), showing a clear increase of $\Delta \varphi$, as the field increases from 6 to 15\,T. This result should be taken into account for a further development of the theoretical model \cite{Petitgrand1999,Plakhty2003}. On the other hand, the limiting case of very high fields predicted by the theory seems to be quite robust. Indeed, it is natural to expect that when the Zeeman energy significantly exceeds the other relevant energy terms, the staggered moment should align precisely perpendicular to the $\mathbf{B}$ and the hysteresis should vanish. Thus, it would be interesting to study the AMR behavior at higher fields to check whether the $\Delta \varphi(B)$ dependence reaches a maximum and turns to a decrease.

Another point which should be better understood is the evolution of the hysteresis with temperature. It is clear that the magnetic energy changes at the transitions between the phases I, II, and III induced by temperature. As mentioned above, this may be a reason for the nonmonotonic temperature dependence of the hysteresis width shown in Fig.\,\ref{fig:Hyst}. However, a more detailed analysis of the data would require a further development of the theory.

The presence of the hysteresis in the AMR of the SC samples, implies that superconductivity and steady AF order coexist at least in the narrow interval $0.12\leq x \leq 0.13$ on the border of the SC doping range. Noteworthy, a weak hysteretic anomaly has also been found in the interlayer magnetoresistance of a SC sample with $x = 0.13$ for an out-of-plane rotation in a strong magnetic field \cite{Kartsovnik2011}. While the exact reason for the latter anomaly needs a separate investigation, it is most likely related to the AF ordered magnetic system.
At present we cannot rule out a slight inhomogeneity of cerium distribution or oxygen defects as possible reasons for the existence of two phases in a crystal. However, the subtle differences in the behavior of $\Delta \varphi$ seen in Figs.\,\ref{fig:Hyst} and \ref{fig:Hyst_25K_60K} hint towards a systematic $x$-dependence of the hysteresis in the present experiment. If this is indeed the case, it would mean that the hysteresis and hence the AF state are intrinsic to each of the present doping levels, including the SC ones.
For clarifying the situation, more thorough magnetization and spectroscopic studies of high-quality SC crystals on the lower edge of the SC doping region would be very interesting.

\section{Conclusion}

We have carried out a detailed study of the angle-dependent magnetoresistance (AMR) of underdoped NCCO crystals, focusing on a narrow doping range around the border of the SC region. On top of a smooth $90^\circ$-periodic AMR background, reflecting the square electronic/magnetic anisotropy of the system, two pronounced features have been observed on the non-SC samples. A sharp feature in the intermediate field range, $1\,\mathrm{T}\leq B\leq 4$\,T, is identified with the field-induced transition from the noncollinear to the collinear AF state as the field is turned away from the Cu--O--Cu direction towards Cu--Cu.
For all samples, at higher fields a remarkable hysteretic anomaly is found in the AMR patterns around the $[100]$ direction. It is accompanied by an irreversible step of magnetoresistance in magnetic-field sweeps performed in a certain sequence. Remarkably, the same high-field anomaly is also found in the normal-state magnetoresistance of the SC samples.

The hysteretic AMR can be qualitatively explained in terms of reorientation of the AF ordered Cu$^{2+}$ spins in the high-field collinear state. The key point in the proposed scenario is that the direction of the staggered moment in this state is not exactly perpendicular to the applied field but is inclined by an angle depending on the field strength, as was shown earlier for undoped $\mathrm{Pr}_{2}\mathrm{CuO}_4$  \cite{Plakhty2003}. This leads to ``freezing'' of the spins in a metastable state when the rotating magnetic field passes through the $[100]$ direction. The spins relax to the stable state as the field is further turned to a high enough angle $\varphi^*$.
The hysteresis can be traced up to temperatures as high as $\sim 100$\,K. This provides a solid support for the AF Cu$^{2+}$ system to be at the origin of the effect, as opposed to the low-temperature antiferromagnetism of Nd$^{3+}$ ions, which was suggested to be responsible for the similar behavior in low-doped NCCO \cite{Wu2008}. Within the existing theoretical model \cite{Plakhty2003,Petitgrand1999} it is still unclear why the hysteresis width would increase with the field strength, as observed in the experiment. Besides further elaborating the theory, it would be interesting to perform the AMR experiment at higher magnetic fields. The present model suggests that the hysteresis should strongly diminish and probably vanish when the Zeeman energy term in Eq.\,(\ref{energy-tot}) becomes much larger than the contribution from the pseudodipolar interaction. Using the experimentally obtained critical fields of the noncollinear -- collinear transition at $\mathbf{B}\| [100]$ and $\mathbf{B}\| [110]$, we estimate that a field of $\sim 30$\,T must be sufficiently high for checking this. As another test of the proposed scenario, one could perform a similar study on underdoped $\mathrm{Pr}_{2-x}\mathrm{Ce}_{x}\mathrm{CuO}_4$. This compound is very similar to NCCO, showing, in particular, the transition between noncollinear and collinear AF states in a magnetic field. However, by contrast to NCCO, it has only one AF configuration at all temperatures \cite{skan93,Sachidanandam1997}. Therefore it is expected to exhibit the same behavior at high fields, but with the hysteresis width monotonically depending on temperature.

Finally, the observation of the hysteretic anomaly in our SC samples indicates that the long-range AF order survives up to the SC doping range, at least at temperatures above $T_c$. However, one needs further purposeful studies on samples with well defined crystal quality in the doping range close to the border of the SC region to understand how the two orders coexist and to eliminate possible spurious effects of chemical inhomogeneity.

\begin{acknowledgments}
This work was supported in part by the Deutsche Forschungsgemeinschaft (DPG) via Grant GR 1132/15. A.D. gratefully acknowledges financial support from Consejo Nacional de Ciencia y Tecnolog{\'i}a (CONACYT). Z.H. thanks the "Lee's Pharmaceutical-Kanya Lee scholarship" for making possible the stay at the Technische Universtit{\"a}t M{\"u}nchen and the Walther Mei{\ss}ner-Institut.
\end{acknowledgments}

\appendix*
\section{Low-temperature field dependence of magnetoresistance at $\varphi = 0^\circ$ and $45^\circ$
}
\label{AppA}
\begin{figure}
  \centering
  \includegraphics[width=0.45\textwidth]{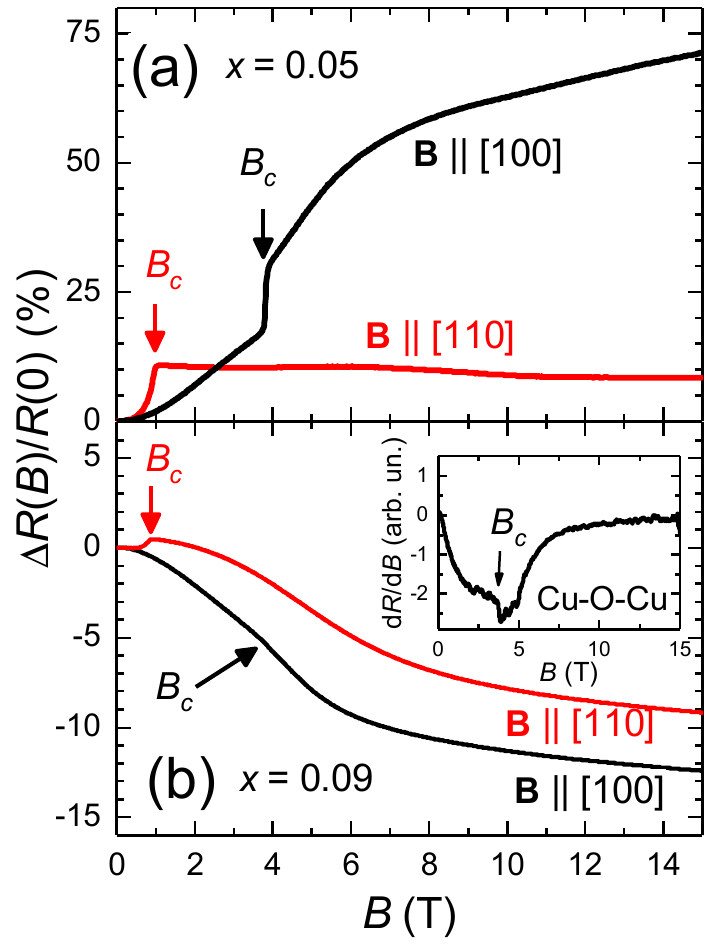}
  \caption{(Color online) Field-dependent magnetoresistance of the non-SC samples with $x = 0.05$ (a) and 0.09 (b), at $T = 1.4\,\mathrm{K}$. The inset in (b) shows the derivative $dR/dB$ for the $x=0.09$ sample in field along $[100]$. $B_c$ is the critical field of the transition between the noncollinear and collinear AF spin configurations.
}
\label{fig:RvsB_N5_N9_lowT}
\end{figure}

Figure\,\ref{fig:RvsB_N5_N9_lowT} shows the low-temperature field dependence of interlayer magnetoresistance of the underdoped non-SC samples, $x=0.05$ and $0.09$, for a magnetic field applied parallel to the $\mathrm{CuO}_2$ layers, in the $[100]$ and $[110]$ directions. For  $\mathbf{B}\|[110]$, both samples show a clear kink at $B_{c}(\varphi=45^\circ)\approx 1\,\mathrm{T}$. This kink indicates the transformation of the antiferromagnetically ordered Cu$^{2+}$ spins between the noncollinear and collinear configurations, like it was observed earlier in lower doped crystals \cite{Lavrov2004,Chen2005,Wu2008}. Similarly, for  $\mathbf{B}\|[100]$, the magnetoresistance of the $x = 0.05$ sample has a jump at $B_{c}(0^\circ) = 3.8\,\mathrm{T}$ due to the first order spin-flop transition between the noncollinear and collinear states. The spin-flop transition is less discernible for the $x = 0.09$ sample. However, the field-derivative of the resistance, shown in the inset in Fig.\,\ref{fig:RvsB_N5_N9_lowT}(b), has a step at $\approx 3.8$\,T which is likely associated with this transition.

For the $x = 0.05$ sample the smooth part of the field-dependent magnetoresistance, shown in Fig.\,\ref{fig:RvsB_N5_N9_lowT}(a) is very different at $\mathbf{B}\|[100]$ and $\mathbf{B}\|[110]$: it has a strong positive slope in the former case and is almost flat, just weakly negative in the latter. A qualitatively similar, though somewhat stronger, anisotropy has been found on lower doped samples, $x = 0.025$ and $0.033$ \cite{Chen2005,Wu2008} as well as on low-doped Pr$_{1.3-x}$La$_{0.7}$Ce$_x$CuO$_4$ \cite{Lavrov2004}.  As doping is increased to $x = 0.09$, the anisotropy inverts and becomes much weaker, see Fig.\,\ref{fig:RvsB_N5_N9_lowT}(b); both $R(B)$ curves have negative slopes and are almost parallel to each other, slowly saturating towards high fields.



\begin{thebibliography}{44}
\bibitem {Comment_sample10}
  A thorough inspection of
  transport properties of the sample represented in Fig.\,\ref{fig:N10_Bstar}
  revealed a likely presence of excessive oxygen typical of insufficiently
  annealed NCCO crystals. This is probably a reason of a relatively sharp
  magnetoresistance kink at the spin-flop transition at $\mathbf{B}\|[100]$
  [cf. the very weak feature in Fig.\,\ref{fig:RvsB_N5_N9_lowT}(b) for a
  similarly doped sample]. Nevertheless we use the detailed dataset obtained on
  this sample for illustrating representative features of the memory
  effect.
  \BibitemShut {Stop}%
\end{thebibliography}

\providecommand{\noopsort}[1]{}\providecommand{\singleletter}[1]{#1}%

\end{document}